# *IEEE* Access

Multidisciplinary ⋅ Rapid Review ⋅ Open Access Journal

## RESEARCH ARTICLE

# Multi-Dimensional Data Compression and Query Processing in Array Databases

## MINSOO KIM⁠, HYUBJIN LEE⁠, AND YON DOHN CHUNG⁠, (Member, IEEE)
Department of Computer Science and Engineering, Korea University, Seongbuk-gu, Seoul 02841, Republic of Korea

Corresponding author: Yon Dohn Chung (ydchung@korea.ac.kr)

This work was supported in part by the National Research Foundation of Korea (NRF) funded by the Korean Government [Ministry of Science and ICT (MIST)] under Grant NRF-2020R1A2C2013286, in part by the Basic Science Research Program through the National Research Foundation of Korea (NRF) funded by the Ministry of Education under Grant NRF-2021R1A6A1A13044830, and in part by the MSIT under the Information and Communications Technology (ICT) Creative Consilience Program Supervised by the Institute for Information & Communications Technology Planning & Evaluation (IITP) under Grant IITP-2022-2020-0-01819.

**ABSTRACT** In recent times, the production of multidimensional data in various domains and their storage in array databases has witnessed a sharp increase; this rapid growth in data volumes necessitates compression in array databases. However, existing compression schemes used in array databases are general-purpose and not designed specifically for the databases. They could degrade query performance with complex analytical tasks, which incur huge computing costs. Thus, a compression scheme that considers the workflow of array databases is required. This study presents a compression scheme, SEACOW, for storing and querying multidimensional array data. The scheme is specially designed to be efficient for both dimension-based and value-based exploration. It considers data access patterns for exploration queries and embeds a synopsis, which can be utilized as an index, in the compressed array. In addition, we implement an array storage system, namely MSDB, to perform experiments. We evaluate query performance on real scientific datasets and compared it with those of existing compression schemes. Finally, our experiments demonstrate that SEACOW provides high compression rates compared to existing compression schemes, and the synopsis improves analytical query processing performance.

## I. INTRODUCTION

Multidimensional data, for example, observation data of the Earth's atmosphere and a colorized planet map from a satellite, are widely used in various domains. Data produced by commercial applications, such as financial, statistical, or stock data, can also be modeled as multidimensional data. These data are frequently stored in files and treated in specialized applications in each domain. With this approach, it is difficult to handle scalability, data versioning, and fast query processing. Conversely, array databases afford fast query processing of large volumes of data and thus have received attention in the scientific domain [1]. SciDB [2], one of the most widely used array databases, supports the storage and management of the telescope data generated by the Large Synoptic Survey Telescope (LSST) project. Similarly, TileDB [3] is used in biomedical research to store genomics data.

Recently, owing to their continuous generation, these data have increased to massive volumes; for instance, the telescope used in the LSST project is expected to generate 20-30 TB of data per night [4]. The large volume of data requires enormous storage space and incurs an I/O bottleneck. Thus, databases require compression to function efficiently [5]. However, thus far, compression is only an auxiliary tool in existing array databases; most of the compression schemes used in the latest array databases are general-purpose [6]. Compression has been studied in domains other than databases. Many data processing domains–including image, video, and 3D objects–use compression schemes specific to their data. In these domains, each scheme is designed only for its individual data type, considering the characteristics of target data. Accordingly, the compression schemes used in databases require this approach as well.

In databases, it is important that the compression scheme balances the query performance and compression efficiency. A compression scheme that incurs high computational costs for a high compression ratio is undesirable as it significantly degrades query performance. Hence, database systems use lightweight compression schemes that are computationally less expensive and simpler [2], [3], [7], [8]. These schemes provide reasonable query performance; however, they do not significantly reduce the storage volume as they have a poor compression ratio.

Several compact array representations providing efficient query processing have been proposed for array databases [6], [9], [10], [11], [12]. For example, COMPASS [6] converts an array into a bin-based index, and $k^2-raster$ [9] adopts a tree-based representation. These methods are specially designed by considering the data access pattern of specific queries. However, array databases have a rich set of operators, which are optimized in the array-oriented representation. To perform other queries, a compressed array should be reorganized as a plain array. Accordingly, queries that can be performed with the newly designed data representations are limited.

In this paper, we propose a compression scheme, **S**ynopsis **E**mbedded **A**rray **Com**pression using **W**avelet Transform (SEACOW). The proposed method is based on the array-oriented representation, which is compatible with existing array databases. In particular, our compression scheme focuses on balancing between high compression ratio and query performance. SEACOW is optimized for both dimension-based exploration and value-based exploration, which are frequently used in array data analysis tasks.

First, SEACOW provides efficient data access for multidimensional data. In multidimensional data, a dimension-based exploration is used for data visualization, cropping, or range selection. These tasks access array cells in a particular region, where the region is defined by a combination of continuous ranges in dimensions. SEACOW retains the relative position of the cells after the compression. As the adjacent cells in the original multidimensional space remain within the vicinity of each other in the compressed data, the random access for the dimension-based exploration can be minimized. In addition, SEACOW supports a partial decompression that retrieves only a specific part of the array. This feature helps to reduce the computational cost of decompressing unnecessary array regions.

Second, SEACOW provides useful features for value-based exploration. Multidimensional data are frequently used in analytical tasks, such as finding stars in the telescope image [13], discovering regions of deforestation in the Amazon rainforest from satellite data [14], and detecting highly pressurized regions in computational fluid dynamics (CFD) simulation results [15]. These analytical tasks are based on value-based exploration that finds a specific value in the data. Generally, these tasks require full scanning and the entire data should be decoded from the compressed format leading to substantial I/O and computational costs for decompression. To overcome this shortcoming, SEACOW embeds a data structure called a synopsis that includes summary data and an index for attribute values. The synopsis could help to approximately recognize the data before the decompression and estimate a candidate region for scanning. It would be helpful to improve the value-based exploration performance.

We also introduce a multidimensional scientific database (MSDB): our prototype implementation. MSDB is designed to store compressed scientific data and perform queries on them. It has several built-in operators such as insertion, range selection, and filter. The storage module of the database supports various array compression schemes, including SEACOW, to compactly store array data. In addition, it enables addition of new compression schemes or query operators as a user-defined function in C++. MSDB provides a C++ API library to access and query the database. It is implemented in C++17, and details of the implementation can be freely accessed at Github.[1]

The contributions of this paper are as follows:

- We propose a lossless array compression method, SEACOW, that supports efficient exploration query processing in a compressed array.
- We introduce a tree index structure, HMMT, that minimizes storage requirements.
- We present algorithms to perform queries on a compressed array using SEACOW.
- We implement an array storage, MSDB, and perform experiments on the system.
- We evaluate the compression and query performance of SEACOW and compare it with well-known compression schemes.

The remainder of this paper is organized as follows. We describe the background and related work in Section II. We present SEACOW in Section III and the query processing algorithm on a compressed array using SEACOW in Section IV. In Section V, we present a performance comparison of SEACOW with other existing methods. Finally, we draw conclusions and discuss future work in Section VI.

## II. BACKGROUND & RELATED WORK
### A. ARRAY DATA MODEL
The array data model is composed of dimensions, cells, and attributes. The dimensions define a multidimensional space to represent the array data, and each dimension has consecutive integers with a range limitation. The $m$-dimensional array has a set of $m$ dimensions $\mathbb{D} = \{d_0, d_1, \ldots, d_{m-1}\}$. For each combination of dimension values, an array cell is present. Accordingly, every array cell has a particular location in a multidimensional space and holds a fixed number of attributes, each of which is identified by its unique name. For simplicity, in the rest of this paper, we proceed on the assumption that the array contains only one attribute.

---

[1] "https://github.com/KUDB/MSDB"

Generally, due to their massive size, multidimensional arrays are partitioned into sub-arrays. There are several partitioning techniques, depending on their levels and whether the sub-arrays are regular [16]. This paper uses chunking to refer to the array partitioning strategy and a chunk to refer to the partitioned sub-arrays. The chunks can overlap at their boundaries; however, considering the storage efficiency, we consider only non-overlapped chunks in this study.

### B. DISCRETE WAVELET TRANSFORM

The discrete wavelet transform (DWT), also known as wavelet decomposition, is a mathematical tool that is widely used in signal processing. In particular, data compression is a well-known application of wavelet transform. It decomposes the source data by successively applying low-pass and high-pass filters. Each process generates two subbands, which contain coefficients from the filtering operations. The coefficients from the low-pass filter are called approximation coefficients, and the coefficients from the high-pass filter are the detail coefficients. The subband of the approximate coefficients contains many significant data, while the subband of the detail coefficients consists of relatively insignificant data. Thus, the data can be compressed considerably by reducing the overall size of the subband of detail coefficients.

There are various kinds of wavelets. In this paper, we use the Haar wavelet, where the low-pass and high-pass filters consist of $\{\frac{1}{\sqrt{2}}, \frac{1}{\sqrt{2}}\}$ and $\{\frac{1}{\sqrt{2}}, -\frac{1}{\sqrt{2}}\}$, respectively. Conceptually, it is the simplest and fastest wavelet. In particular, the key property of the Haar wavelet is that it generates down-scaled data during the process. According to the low-pass filter, the approximate coefficients are computed by averaging the non-overlapped pairwise values in the Haar wavelet. Then, the subband of the approximate coefficients can be viewed as a summary of the data. In this paper, the subband is used to improve query performance.

### C. RELATED WORK
#### 1) DATA COMPRESSION
There are many compression schemes for multimedia data. JPEG [17], a widely used image compression standard, is based on the Fourier transform. On the other hand, several image compression schemes based on the wavelet transform have also been proposed. EZW [18] is an early study using the wavelet transform in image compression. Subsequently, SPIHT [19] and EBCOT [20] improved the compression performance. However, these techniques are optimized to perform on a two-dimensional array with one-byte integer values. Extending them to $n$-dimensional data would be difficult, and query performance involving heavy computational tasks might be degraded.

Alternatively, data compression has also been studied in scientific domains. SZ [21], [22] is an error-bounded lossy compression scheme for multidimensional scientific data. It serializes a multidimensional array into a one-dimensional array, and predicts successive bits with its fitting curve model. There are rich derived algorithms depending on the type of compressing data or use cases. TTHRESH [23] was a state-of-the-art lossy compression method that is based on the tensor decomposition. It achieves a high compression ratio with low error tolerances. Nevertheless, both techniques seem inappropriate for use in databases as they lead to information loss in the compression process.

Although long studied in various domains, data compression is particularly important in database systems. Previous studies [5], [24], [25] demonstrated that data compression not only reduces storage requirements but also improves query performance. Accordingly, existing databases provide numerous compression schemes in their storage systems. For lightweight compression methods, several fixed-length encoding techniques, such as PFOR, PFOR-DELTA, and PDICT, have been introduced [26]. They provide competitive compression ratios with low computational costs. Bounce [27] introduces a memory-efficient single instruction multiple data (SIMD) approach for bit-packing compression. Conversely, Lempel-Ziv [28], [29] and Huffman coding [30] use variable-length coding. They convert the fixed size of a symbol to the variable size of code words. Both compression schemes are the basis of various compression schemes. For example, they are used in well-known file compression formats, such as Zip, Gzip, and 7z. On the other hand, CodecDB [8] is an encoding-aware columnar database that has a tightly coupled design comprising a data encoding scheme and query engine. In this system, the embedded encoding selector chooses a proper compression scheme for the data. A recent study [31] applies a deep neural network to relational data compression. Compression schemes to reduce the size of binary relations have also been studied [32], [33]. These are based on $k^2-tree$ [34], a data structure resembling a quad-tree.

Queryable compression schemes for raster data suitable for array databases have been studied, where the raster data are the specific forms of multi-dimensional arrays consisting of rows and columns. In particular, the $k^2-raster$ [10] based on $k^2-tree$ supports the raster data containing various integer values. It recursively partitions the array until all the cells in the partitioned array have the same value. In addition, the 2D1D-map and 3D2D-map [11], [12] are also based on $k^2-tree$ and support several spatial queries, such as point access, window, range, and value range queries. However, these schemes only target the two-dimensional array (or raster data). On the other hand, COMPASS [6] reorganizes array cells into a value-indexed representation and makes the array compact. Value-based exploration can benefit from this representation. However, it is inadequate for queries with spatial access, such as dimension-based exploration and grid aggregation [35]. ZFP [36] compressed multidimensional floating-point data with block transform and embedded coding. It can be both lossy and lossless. However, array databases used in the real world still embed only several general-purpose compression schemes. SciDB [2], [13] uses run-length encoding with null-suppression. TileDB [3] has conventional file compression schemes, such as lz4, gzip, and bzip2.

### 2) SYNOPSIS
A synopsis is a small summary of the data and is frequently used to improve query processing performance. Multi-resolution aggregate tree (MRA-tree) [37] is a multi-dimensional index structure that contains several synopses; such as min, max, sum, and count. It can be updated and progressively answer aggregate queries. Similarly, Searchlight [38] utilizes multiple synopses with constraint programming. It first obtains the approximate answer from the synopses and performs a search query over large multidimensional data. Holding various types of synopses is useful for performing aggregate queries. However, each synopsis occupies additional memory and storage space. In contrast, we focus on improving exploration query performance only using a restricted type and size of synopsis. Particularly, we include the synopsis in the overall size of the compressed array.

Furthermore, the wavelet synopsis (or synopsis array) is often used for approximate query processing. A wavelet synopsis is an approximation of the original data produced during the wavelet transform. Chakrabarti et al. [39] and Garofalakis et al. [40] provided general-purpose approximate query processing algorithms with wavelet synopses. Sacharidis et al. [41] proposed a compression scheme for wavelet synopses. However, they targeted only the approximate query processing and focused on minimizing the error of the approximation, rather than targeting the exact query processing. Jahangiri et al. [42] used a wavelet synopsis to range-group-by queries. It supports both approximate and exact query processing algorithms. However, it is specialized only for dimension-based queries. ProDA [43] is an OLAP system that employs wavelets to manage massive data and support various types of queries.

### 3) SCIENTIFIC DATA STORAGE
Scientific data storage systems deal with massive data. In these systems, the data indexing plays an important role in providing efficient data services, such as search and management [44], [45]. MOSIQS [46] utilizes a tree based index data structure called GSM, which is designed for scientific datasets. SciSpace [47] supports metadata indexing for geographically dispersed scientific data files. These systems assume that the scientific data are stored in numerous files with diverse scientific data formats, such as HDF5 and NetCDF. They primarily target indexing the metadata of the files, not the raw scientific data itself. Our work enables the exploration of massive multidimensional data. We embed an index structure and synopsis in the compressed data file. Then, the file tightly integrates into our database system.

**TABLE 1.** Notation table.

| Variables | Description |
|---|---|
| $\mathbb{A}, A'$ | Raw array, sub-array of $\mathbb{A}$ |
| $\mathbb{D}$ | Set of dimensions of an array |
| $d_i$ | $i^{th}$ dimension in $\mathbb{D}$ |
| $C_i, B_j$ | $i^{th}$ chunk, $j^{th}$ block in each chunk |
| $R_i^{est}, R_i^{real}$ | Estimated and exact values of the bit lengths to encode each element in block $B_i$ |
| $\Delta_i$ | Difference between $R_i^{est}$ and $R_i^{real}$ |
| $N_{root}$ | Root node of HMMT |
| $N_i$ | Child node of HMMT whose id=$i$ |
| $N_i.j$ | $j^{th}$ child node of $N_i$ |
| $N_i^{min}, N_i^{max}$ | Min and max element in the fragment of $N_i$ |
| $N_i^{order}$ | Order of significant bit used in $N_i$ |
| $N_i^{width}$ | Required bits to encode each variable of $N_i$ |
| $L_{chunk}, L_{block}$ | Level of HMMT where the node's fragment matches the size of the chunk and block, respectively |
| $\mathcal{B}_i(x)$ | $i^{th}$ significant bit of $x$ |
| $\mathcal{S}_i(x)$ | $i^{th}$ significant bit of $x$ in sign and magnitude representation |
| $\mathcal{R}(A') \mapsto N_i$ | Sub-array $A'$ and node $N_i$ have relationship |
| $\mathbf{PARENT}(N_i)$ | Returns a parent node id of $N_i$ |
| $\mathbf{CHILD}(N_i)$ | Returns a set of child node ids of $N_i$ |

## III. SEACOW: THE PROPOSED METHOD
### A. OVERVIEW
In this section, we illustrate our lossless array compression scheme, namely SEACOW. It compresses a multidimensional array with a high compression ratio while providing efficient exploration and analysis of query performance. Especially, fast query processing can be performed with two embedded data structures: Hierarchical Min-max Tree (HMMT) and a synopsis array. The HMMT hierarchically represents the range of values in each sub-array and is helpful for improving searching query performance while restricting the exploration region. In addition, the synopsis array contains aggregated data from the source array. We can utilize the pre-aggregated array to boost various types of aggregate query performance. These data structures are part of a compressed array and can be retrieved independently. For example, the HMMT is stored in a header of the compressed array, and the synopsis array is written in front of the compressed bit-stream of each sub-array. Accordingly, they can be quickly retrieved without decoding the entire array; moreover, HMMT helps to briefly recognize the source array in advance.

An overview of the SEACOW compression process is presented in Figure 1. First, in the preprocessing step, we build the HMMT and convert the source array into a wavelet transformed array. Then, in the bit-packing step, we pack each sub-array with a minimum length of bits utilizing the HMMT built in the previous step. Finally, we compress the bit-stream using both run-length encoding and Huffman coding in the bit-reduction step. The last step is optional. In the following sections, we describe the components of SEACOW and compression steps in detail. Table 1 presents the notations used in this paper.

### B. ARRAY CHUNKING
In SEACOW, an array is vertically partitioned by attributes, which is similar to the columnar storage of relational databases. Then, an array of a single attribute is chunked into several small sub-arrays. In this division, SEACOW adapts two-level regular chunking, also known as regular-regular chunking [16]. The two-level regular chunking first partitions an array into fixed-size chunks and divides the chunks again into fixed-size blocks. In the chunking strategy, the chunk is the unit of I/O, and the block is the unit of query processing. Accordingly, each chunk is compressed individually and written on the disk, and vice versa. It is noteworthy that with some specific compression methods that support partial decoding, only a part of the chunk can be effectively restored into a plain sub-array. In this case, it is not necessary to decompress all the blocks in the chunk retrieved from the disk; however, only a certain number of blocks can be decompressed. Figure 2 shows an example of a two-level chunking with a two-dimensional 8 × 8 array.

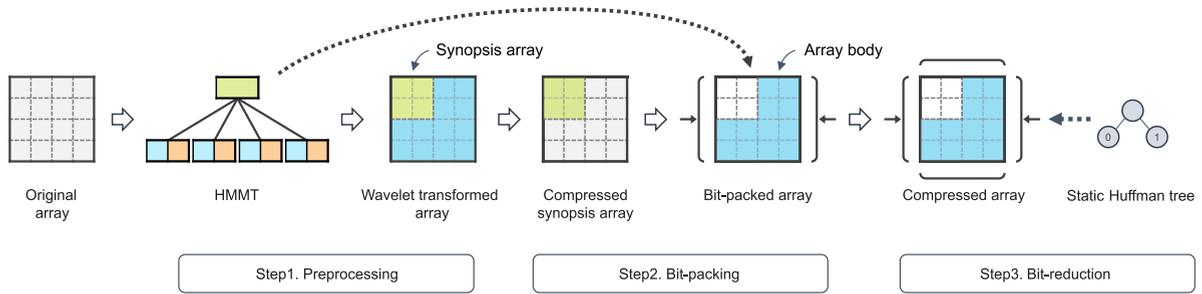

**FIGURE 1.** Overview of SEACOW compression process. The compression involves three steps: preprocessing, bit-packing, and bit-reduction.

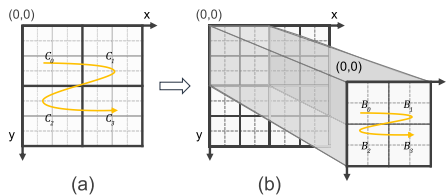

**FIGURE 2.** Example of a two-level chunking. The array in (a) splits into four 4 × 4 chunks, such as $C_0$, $C_1$, $C_2$, and $C_3$. Then, the chunk $C_0$ is further divided into four 2 × 2 blocks, such as $B_0$, $B_1$, $B_2$, and $B_3$, as shown in (b). The chunks and blocks are numbered in row-major order. The arrows over the array show the order.

SEACOW takes two parameters for chunking: a chunk size and a wavelet level, where the wavelet level is also used in the compression process. The size of the first level partitioned sub-array, chunk size, is specified by the user. The block size depends on both the chunk size and the wavelet level, and it is the same as the size of the smallest subband in the wavelet transformed chunk. Accordingly, the higher wavelet level makes a block smaller. In Section V-A, we describe the mechanism of finding the suboptimal parameters for several datasets.

### C. SYNOPSIS ARRAY

The synopsis array is a summary of the source array, which is downsampled by average pooling. This array is generated as a result of the wavelet transformation and is also called wavelet synopsis. In the wavelet transform, an array is divided into a set of approximate coefficients and several sets of detail coefficients. In a Haar wavelet transform, which is used in SEACOW, the approximate coefficients are obtained by averaging the elements of the neighboring cells. As a result, the transformed result is identical to the result of the grid aggregation that splits the source array into non-overlapped sub-arrays and applies an average aggregate function to values in the sub-array. In this paper, we call the set of approximate coefficients the synopsis array.

The synopsis is used to improve query processing performance. In the scientific data domain, a common task explores multidimensional data with interactive visualization tools [48]. The task primarily begins with a thumbnail, which is a scaled-down version of the source array, and requires a heavy aggregate query over the entire array cells. Thus, by providing the synopsis array, we can instantly serve the summarized array without additional query processing. Furthermore, it is also beneficial to process various types of aggregate queries such as summing or averaging values in a sub-array.

### D. HMMT

The HMMT (hierarchical min-max tree) is a hierarchical data structure that represents the value range of each sub-array hold. There are several HMMTs according to the number of attributes in an array. The tree is primarily used for compressing each attribute array and can also be extended as an index for value-based exploration queries. HMMTs are compressed with the array and stored on a disk. In the HMMT compression process, we utilize a hierarchical relationship of the tree to minimize storage requirements. Thus, we can build a fine-grained index structure for an array with lightweight volume. This is important to maintain the high compression ratio of the overall data.

#### 1) TREE DESIGN

HMMT is a tree index based on multidimensional space decomposition. The appearance of the tree is similar to many existing trees, however, it is not fixed to a particular $m$-dimensional domain and can be multidimensional according to the source data. For example, it can be a quadtree for two-dimensional data and an octree for three-dimensional data. Figure 3 shows an example of the HMMT. The root node is denoted as $N_{root}$, and the others are denoted as $N_i$, where $i$ denotes its lineage, e.g., $N_{i.j}$ means that the node has a parent node $N_i$ and is the $j$th child of the parent node. All intermediate nodes of the $m$-dimensional HMMT have $2^m$ number of child nodes. Then, we define a function $CHILD(N_i)$ that returns a set of child node ids of $N_i$. For example, in a two-dimensional array, $N_i$ has four child nodes: $N_{i.0}$, $N_{i.1}$, $N_{i.2}$, and $N_{i.3}$, where $CHILD(N_i) = \{0, 1, 2, 3\}$.

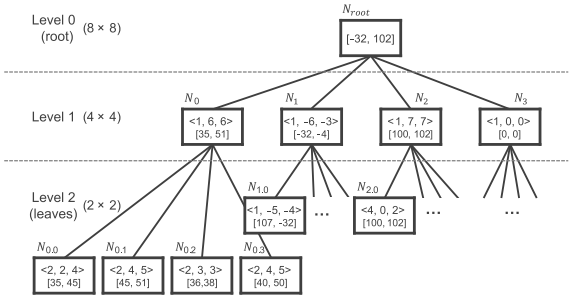

**FIGURE 3.** Example of three-level HMMT. In each node, its tuple ⟨*order*, $\mathcal{S}_{\text{order}}(min)$, $\mathcal{S}_{\text{order}}(max)$⟩ is written at the first row, and the actual values before the conversion [*min, max*] are listed at the bottom. The root node has only *min* and *max*.

On each tree level, the array is logically partitioned into non-overlapping sub-arrays called *fragments*. It is not a physical partition, such as chunks or blocks, and only represents each node's region of interest. In a $m$-dimensional tree, a parent fragment is decomposed into two on each dimension, and partitioned into $2^m$ number of child fragments. For example, in a two-dimensional array, a fragment is recursively subdivided into $2 \times 2$ child fragments, while, in a three-dimensional array, it is subdivided into $2 \times 2 \times 2$ child fragments. Let a function $\mathcal{R}$ indicates the node $N_i$ having an identical fragment region as the sub-array $A'$ and is denoted as follows: $\mathcal{R}(A') \mapsto N_i$.

The node $N_i$ knows the minimum element ($N_i^{\min}$) and maximum elements ($N_i^{\max}$) of the $A'$. Accordingly, all elements in $A'$ exist between $N_i^{\min}$ and $N_i^{\max}$. For simplicity, these variables are also called *min* and *max* of $N_i$ and are defined as follows:

$$N_i^{\min} = \min\left(\{x \mid \forall x \in A'\}\right)$$
$$N_i^{\max} = \max\left(\{x \mid \forall x \in A'\}\right)$$

In the multidimensional domain, the number of nodes increases rapidly as the level of HMMT increases. The storage requirement of a tree could be exceedingly large if HMMT is built to a very fine-grained level. To keep the tree lightweight, the tree nodes treat each variable in binary form and hold only its significant bits. Although the subsequent bits are discarded, the HMMT is sufficient to perform its roles. First, as the main role of HMMT, it informs the estimated length of bits for bit-packing sub-arrays in the array compression process. This can be performed using only the significant bit of the *min* and *max* for each sub-array. In addition, HMMT can serve as an index structure. Because the variables are hierarchically expressed starting from the root node, the *min* and *max* values estimated from the significant bits are sufficiently close to the actual value as it goes down to lower-level nodes.

In this conversion, we use the sign and magnitude representation for negative values. First, let $\mathcal{B}_n(x)$ be $n$th significant bits of $x$ and $\mathcal{B}_n(x)$ be the zero if there is no $n$th significant bits for $x$. Then, we define $\mathcal{S}_n(x)$ that indicates the sign of $x$ with $n$th significant bits of its magnitude value. It is defined as follows:

$$\mathcal{S}_n(x) = \text{SIGN}(x) \times \mathcal{B}_n(|x|),$$

where **SIGN**$(x)$ returns 1 for $x \geq 0$ and $-1$ for $x < 0$. In the rest of the paper, we just call $\mathcal{S}_n(x)$ as a $n$th significant bit of value $x$. We also omit $n$ from $\mathcal{B}_n(x)$ and $\mathcal{S}_n(x)$ when $n=1$ for simplicity of exposition.

We transform the variables of the tree nodes into significant bits using the function $\mathcal{S}_n(x)$. As a result, a set of variables ⟨*min, max*⟩ is converted to a tuple ⟨*order*, $\mathcal{S}_{\text{order}}(min)$, $\mathcal{S}_{\text{order}}(max)$⟩ in each node, where *order* is the sequence of the significant bit referenced by the node. Meanwhile, $N_{root}$ still holds *min* and *max* to provide the exact value range of the array. In the following sections, we describe the tree-building process and the compression algorithm. The detailed process for varying the *order* is also described in Section III-D3.

#### 2) TREE ORGANIZING
The tree organizing process starts from the leaf nodes, the number of which was set as the number of blocks in the array. The boundary of the fragment region of each node is aligned to the same as the physical partition of the array. This means that each leaf node has a one-to-one relationship with a specific block in the array. Accordingly, the leaf level is also called as a *block level* and denoted as $L_{\text{block}}$. In $L_{\text{block}}$, each node $N_i$ has the minimum and maximum element in its correlated block $B_j$, where $\mathcal{R}(B_j) \mapsto N_i$. A simple way to find the $N_i^{\min}$ and $N_i^{\max}$ is by traversing all the elements in $B_j$.

Building an intermediate level is similar to the constructing process of the existing trees based on spatial decomposition. In the intermediate levels, the child nodes are grouped according to the position of their fragments; two neighboring child nodes are grouped for each dimension. In general, intermediate nodes have the same number of child nodes, e.g., four child nodes in a two-dimensional tree, and eight child nodes in a three-dimensional tree. The fragment of the intermediate node is identical to the region where the fragments of its child nodes are merged. The intermediate nodes can find a minimum element and maximum element in its fragment while taking the *min* and *max* from its child nodes. During the process, when the nodes satisfy $\mathcal{R}(C_i) \mapsto N_i$, the level is called as *chunk level* and denoted as $L_{\text{chunk}}$. The building on a new intermediate level ends when the remaining child nodes are insufficient to create a group. If merging is no longer available as there is only one child node in any dimension, all nodes in the latest level are grouped to create the root node $N_{root}$. The $N_{root}$ covers the entire array region, and now $N_{root}$ knows the minimum and maximum elements of the array.

---

**Algorithm 1:** HMMT Compression Algorithm

**Input** : an HMMT for an array whose data type is $Ty$
**Output**: a compressed bit-stream of the HMMT

1 /* Initialize */
2 Initialize output stream *out*
3 $\mathbb{N}_{root}^{order} \leftarrow 1$
4 $\mathbb{N}_{root}^{width} \leftarrow sizeof(Ty) \times 8$
5 $\mathbb{N}_{node} \leftarrow \{N_{root}\}$
6
7 **while** $\mathbb{N}_{node} \neq \emptyset$ **do**
8     $N_i \leftarrow \mathbb{N}_{node}.pop\_front()$
9     $\mathbb{N}_{child} \leftarrow \{N_{i,j} \mid \forall j \in CHILD(N_i)\}$
10     $n \leftarrow N_i^{order}$
11     *finished* $\leftarrow$ *false*
12
13     /* Encode current node */
14     *out* $\ll$ setw($N_i^{width}$)
15     **if** $N_i = N_{root}$ **then** // For root node
16        *out* $\ll N_{root}^{min} \ll N_{root}^{max}$
17     **else** // For child nodes
18        $p \leftarrow PARENT(N_i)$
19        **if** $N_p^{order} = N_i^{order}$ **then**
20           $\Delta_{min} = \mathcal{S}_n(N_i^{min}) - \mathcal{S}_n(N_i^{min})$
21           $\Delta_{max} = \mathcal{S}_n(N_p^{max}) - \mathcal{S}_n(N_i^{max})$
22           *out* $\ll \Delta_{min} \ll \Delta_{max}$
23        **else**
24           /* *order* is updated */
25           *out* $\ll |\mathcal{S}_n(N_i^{min})| \ll |\mathcal{S}_n(N_i^{max})|$
26        **end**
27     **end**
28
29     /* Update child nodes */
30     **if** $\mathcal{S}_n(N_i^{min}) = 0$ *and* $\mathcal{S}_n(N_i^{max}) = 0$ **then**
31        *finished* $\leftarrow$ *true*
32     **else if** $\mathcal{S}_n(N_i^{min}) \neq \mathcal{S}_n(N_i^{max})$ **then**
33        // Pass current *order*
34        **foreach** $N_{i,j}$ in $\mathbb{N}_{child}$ **do**
35           $N_{i,j}^{order} \leftarrow N_i^{order}$
36           $N_{i,j}^{width} \leftarrow \mathcal{B}(\mathcal{S}_n(N_i^{max}) - \mathcal{S}_n(N_i^{min}))$
37        **end**
38     **else** // Pass next *order*
39        $m, end \leftarrow findChildNodeOrder(N_i)$
40        *finished* $\leftarrow updateChilds(N_i, m, end, out)$
41        Find *jump* and write it to *out*
42        *out* $\ll$ setw(1)$\ll 0 \times 1$ // set end bit
43     **end**
44
45     /* Add child nodes to list if further encoding is required */
46     **if** *not finished* **then**
47        $\mathbb{N}_{node} \leftarrow \mathbb{N}_{node} \cup \mathbb{N}_{child}$
48     **end**
49 **end**
50 **return** *out*

---

**Algorithm 2:** Algorithm for Finding and Updating *Order* Variable for Child Node

**Input** : node $N_i$
**Output**: next order $n$ and end pos of jump *end*

1 **Function** findChildNodeOrder($N_i$):
2     $n \leftarrow N_i^{order}$
3     **repeat**
4        $n \leftarrow m + 1$
5        $sMin' \leftarrow \mathcal{B}_n(|N_i^{min}|)$
6        $sMax' \leftarrow \mathcal{B}_n(|N_i^{max}|)$
7     **until** $sMin' = sMax'$ *and* $sMax' \neq 0$
8     $end \leftarrow max(sMin', sMax')$
9     **return** $\{n, end\}$
10 **end**

**Input** : node $N_i$, next order $m$, end pos of jump *end*, output stream *out*
**Output**: flag indicating whether child nodes are needed to encode *finished*

11 **Function** updateChilds($N_i, m, end, out$):
12     *finished* $\leftarrow$ *false*
13     $n \leftarrow N_i^{width}$
14     *out* $\ll$ setw($|\mathcal{S}_n(N_i^{max})|$)
15     **if** $\mathcal{S}_m(N_i^{max}) \neq 0$ **then**
16        *out* $\ll m - n$
17        Set *order* of its child nodes to $m$
18        Set *width* of its child nodes to $\mathcal{B}(end)$
19     **else**
20        *out* $\ll 0$
21        *finished* $\leftarrow$ *true*
22     **end**
23     **return** *finished*
24 **end**

---

### 3) TREE COMPRESSION

In this section, we present the lossy compression algorithm for an HMMT. In the compression process, the algorithm exploits the hierarchical relationship of parent-child nodes to encode the tree. The *min* and *max* variables in tree nodes have the following relationship between the parent-child nodes.

$$N_i^{min} \leq N_{i,j}^{min} \leq N_{i,j,k}^{min}$$
$$N_i^{max} \geq N_{i,j}^{max} \geq N_{i,j,k}^{max}$$

Similarly, the significant bits of the variables also have the same relationship; $\mathcal{S}_n(min)$ is constantly increasing and $\mathcal{S}_n(max)$ is decreasing.

Algorithm 1 presents the compression process for an HMMT. We assume that the given HMMT is built for an attribute array whose data type $Ty$ has $sizeof(Ty) \times 8$ bits length. The compressed result is written in output bit-stream *out*. The bit-stream has a function $setw(x)$ that adjusts an output bit length to $x$ bits. We also show an example tree for an array of single-byte integers in Figure 4.

The algorithm is executed in a top-down fashion starting from the $N_{root}$. We temporarily assign one more variable *width* in each node, where *width* is the number of required bits to encode each variable of the node. In the root node, *order* and *width* are initialized to one and $sizeof(Ty) \times 8$ bits, respectively. We encode the root node (Lines 15-16). Only the variables *min* and *max* of the root node are encoded in their real values.

Then, we update its child nodes, including *order* and *width*. As we mentioned previously, *order* is the sequence of the significant bit referenced by the node. When encoding a child node, only $n^{th}$ significant bits of *min* and *max* are recorded, not the exact values. This allows the tree to be compressed considerably.

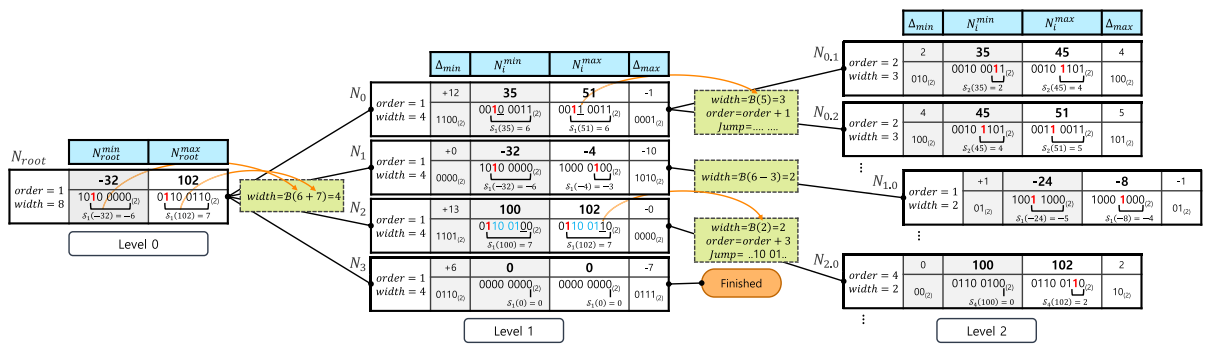

**FIGURE 4.** Example tree composition in HMMT compression process. In the compression process, each node has a set of four variables, such as *order*, *width*, $N_i^{min}$, and $N_i^{max}$). The HMMT compression utilizes the hierarchical relationship of parent-child nodes. In each node, *order* and *width* are updated by its parent node. Then, by comparing the variables of the node with those of the parent node, $\Delta_{min}$ and $\Delta_{max}$ are calculated and written to the compressed bit-stream. The detailed compression process is depicted in Algorithm 1.

The *order* is determined by the parent node. There are three cases as follows. (1) If there are no further significant bits in both *min* and *max*, we set the flag *finished* and ignore its child nodes (Line 30-31). (2) The current *order* is handed over to the child nodes if the significant bits of *min* and *max* are different (Line 32-37). For example, in Figure 4, $N_{root}$ has *order*=1, and the first significant bits of *min* and *max* are different, where $\mathcal{S}_1(-32)=-6$ and $\mathcal{S}_1(102)=7$. Then, its child nodes ($N_0$-$N_3$) inherit *order* from $N_{root}$ and also encode the first significant bits of their *min* and *max*. (3) If the significant bits are identical in both *min* and *max*, *order* is increased to move on to the next significant bit (Line 38-42). In this case, first, we find the next *order* value for its child nodes. The detailed algorithm for finding the next *order* is illustrated in a function *findChildNodeOrder* of Algorithm 2 (Line 1-10). In the function *findChildNodeOrder*, it compares whether the $n^{th}$ significant bits of *min* and *max* of the node are identical. Second, we update the child nodes with the newly found *order* value. The updating is performed in a function *updateChilds*, which is depicted in Algorithm 2 (Line 11-24). The function returns a flag indicating whether the child nodes should be further encoded. Finally, in the current node, we examine the *jump* and write it in the output stream, where *jump* is a set of consecutive identical bits starting from the current significant bit of the *min* and *max*.

Let us suppose the node $N_2$ in Figure 4 is currently being encoded, where *order*=1. Then, any *x* in the fragment of $N_2$ satisfies $100 \leq x \leq 102$. Accordingly, it also satisfies $\mathcal{S}(100) \leq \mathcal{S}(x) \leq \mathcal{S}(102)$. This means that the first significant bits of *min* and *max* in the child nodes have no further change. Now, the algorithm increases *order* for the child nodes and moves on to more precise values. In this process, the increment of *order* can be more than one. Let $N_i^{min}=100$ $(0110\ 0100_{(2)})$, $N_i^{max}=102$ $(0110\ 0110_{(2)})$, and *order*=1. The two variables are close to each other, and they satisfy $\mathcal{B}_2(|100|)==\mathcal{B}_2(|102|)$ and $\mathcal{B}_3(|100|)==\mathcal{B}_3(|102|)$;

however, their fourth significant bits are different. This means that all bits between the first significant bit and the fourth significant bit are the same. These bits (..10 01..) are written in *jump*, and the *order* of child nodes is increased to four. In Figure 4, *jump* are colored in blue in $N_2$.

Now, encoding $N_i$ is finished, and its child nodes are added in the $L_{node}$ to be encoded (Line 46-48). Then, the algorithm repeats the encoding for all nodes in list $\mathbb{N}_{node}$. Except for the $N_{root}$, the other nodes are treated as child nodes, and only the significant bits of their *min* and *max* are considered. In this process, if the *order* is inherited from the parent node, we use the delta, which indicates the difference of the significant bits between the current node and its parent node (Line 17-26). The changes of *min* and *max* variables of a node exist only in one direction, either continuously decreasing or increasing. Accordingly, $\mathcal{S}_n(min)$ and $\mathcal{S}_n(max)$ follow the same trend, and their sign bit is not required. This process saves a few more bits and increases the compression ratio. On the contrary, if *order* is changed in the current node, we encode $|\mathcal{S}_n(N_i^{min})|$ and $|\mathcal{S}_n(N_i^{max})|$, where *n* is newly assigned *order*.

### E. ARRAY COMPRESSION
In this section, we describe the SEACOW compression process.

#### 1) STEP1: PREPROCESSING
The preprocessing step for the array compression has three main functions: (1) building an HMMT from the source array (described in Section III-D2), (2) applying the wavelet transform on the array, and (3) redefining the blocks. HMMT provides an estimation of the required bits to encode each part of the array and is important for the array compression process. Because the HMMT can be made from the source array, it should be built before the wavelet transform.

When we apply the wavelet transform to the array, the transformation is performed in the chunk unit, for which we use the Haar wavelet. During the process, the elements are split into several wavelet coefficients and the boundary of the block collapses. Therefore, we redraw the block boundaries on each wavelet-transformed chunk. The newly made block parallels the previous one, although it has a different region of interest. Particularly, shuffling of the element during the wavelet transform has specific rules. Let us suppose that there is an $8 \times 8$ array $\mathbb{A}$ (in Figure 5(a)) with its two-level wavelet-transformed array (in Figure 5(a)), where both arrays are composed of a single chunk. Each chunk is partitioned into 16 blocks, as shown in Figure 5(b) and (d). In the figures, the dashed line represents the boundary of the block, and the solid line represents the boundary of the subbands after the wavelet transform is applied. Then, we indicate the relationship between a block and the node in the HMMT (in Figure 5(e)). For example, the blocks in a plain array (in Figure 5(d)) are mapped to the nodes on the $L_{block}$ that have the same size and position as the blocks. However, the blocks in the wavelet transformed array (in Figure 5(b)) are related to the nodes from various levels in HMMT. The related node level depends on the wavelet level to which the block belongs. The relationship can be inferred from the spatial orientation tree of the wavelet transformed array, as shown in Figure 5(a). For example, $B'_7$, $B'_{13}$, and $B'_{15}$ contain some of the detail coefficients, where $wavelet\_level=1$. The source of the coefficients is a set of blocks including $B_{10}$, $B_{11}$, $B_{14}$, and $B_{15}$ in Figure 5(d). Furthermore, $B'_1$, $B'_4$, and $B'_5$ also contain detail coefficients from the set of blocks, and $B'_0$ contains approximate coefficients from them, where $wavelet\_level=2$. Consequently, the blocks in $wavelet\_level=1$ are affected by all blocks of the source array $\mathbb{A}$ from $B_0$ to $B_{15}$. Accordingly, $\mathcal{R}(\{B'_7, B'_{13}, B'_{15}\}) \mapsto N_3$ and $\mathcal{R}(\{B'_0, B'_1, B'_4, B'_5\}) \mapsto N_{root}$.

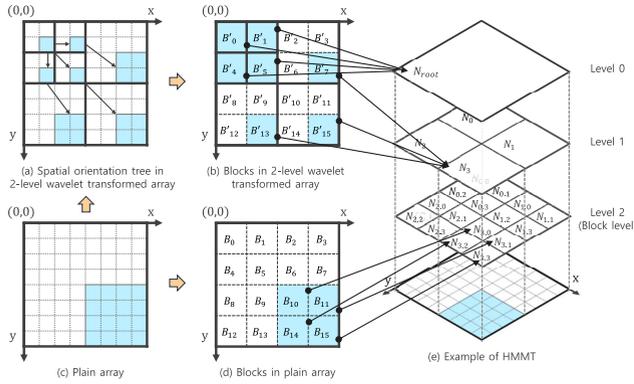

**FIGURE 5.** Example of relationship between blocks and HMMT nodes. In (d), each block of the plain array has a one-to-one relationship with a node located at the block level of HMMT. Conversely, in (b), each block of the wavelet transformed array has a one-to-many relationship with a node. Furthermore, the referred nodes are distributed at different levels. The relationship can be derived from the spatial orientation tree in (a).

#### 2) STEP2: BIT-PACKING
In the bit-packing step, we convert the wavelet-transformed array into a compressed bit-stream. As the chunk is the I/O unit, each chunk is independently compressed and decompressed. The chunks are also split into blocks, and the encoding is performed in the order of the blocks in each chunk. The elements of each block are bit-packed without any information loss. We denote the minimum length of bits required to encode the elements in $B_i$ as $R_i^{real}$. It can be found by examining all cells of the block. It is as follows:

$$R_i^{real} = \max \left( \{\mathcal{B}(|x|) \mid \forall x \in B_i\} \right) + 1,$$

where one bit is added to the sign bit at the end of the equation. Then, each element in $B_i$ is tightly bit-packed using $R_i^{real}$ bits. Conversely, to unpack the block elements, we should know the $R_i^{real}$, which indicates how many bits each element occupies.

On the other hand, because the HMMT has the minimum and maximum element of each block, $R_i^{real}$ can be inferred utilizing the HMMT. The estimated bit length for bit-packing $B_i$ is denoted as $R_i^{est}$. In the header of each compressed block, different $\Delta_i = R_i^{est} - R_i^{real}$ is recorded. Then, $R_i^{real}$ can be deduced from $R_i^{est}$ and $\Delta_i$, which are carried in the HMMT and the block header, respectively.

Here, we describe the process of estimating $R_i^{est}$ from the HMMT. Particularly, each chunk has two types of blocks, and the estimation depends on the type of block. In each chunk, the first block ($B_0$) is a synopsis block that contains approximate coefficients of a wavelet transformed array. Suppose that $B_0$ correlates with $N_j$. Then, any of the approximate coefficients $x \in B_0$ are $N_j^{min} \leq x \leq N_j^{max}$. Accordingly, $R_0^{est}$ for $B_0$ can be obtained as follows:

$$R_0^{est} = \max \left\{ \mathcal{B}\left(\left|N_j^{min}\right|\right), \mathcal{B}\left(\left|N_j^{max}\right|\right) \right\} + 1$$

The rest of the blocks of the chunk contain detail coefficients. Through the wavelet transformation, they might be relatively small, close to zero, compared to the original element. Suppose that the block $B_i$ correlated with $N_k$. We then approximately assume that the detailed coefficients in the block decrease by a factor of two as the wavelet level to which the block belongs increases. Accordingly, the $R_i^{est}$ for $B_i$ is as follows:

$$R_i^{est} = \max \left\{ \max \left\{ \mathcal{B}\left(\left|N_k^{min}\right|\right), \mathcal{B}\left(|N_k^{max}|\right) \right\} + 1 - L, 0 \right\},$$

where $L$ is the wavelet level to which $B_i$ belongs.

Then, each element in $B_i$ is now bit-packed using $R_i^{real}$ bits length.

#### 3) STEP3: BIT-REDUCTION
The bit-reduction is an optional step used under a specific condition, where the dataset contains many duplicated values. The previous compression step, bit-packing, uses a fixed-length code to encode the elements in a block; due to this, all elements in the same block consume a fixed number of $R_i^{real}$ bits. However, it could be inefficient if there is an outlier, the value of which significantly increases $R_i^{real}$. Thus, variable-length coding could be a more effective option. The SEACOW adapts both run-length encoding (RLE) and Huffman coding [30], which is known to be less computational than arithmetic coding. The RLE first encodes consecutive redundant values into one code word. Then, Huffman coding converts a frequently used value (also called symbol) into a short code word, while the less common value is converted into a long code word. The relationship between the value and code word is written in the Huffman table.

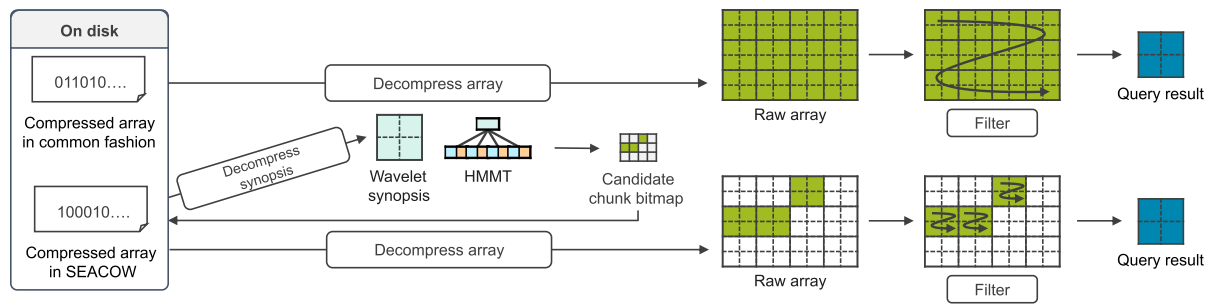

**FIGURE 6.** Overview of the filter query processing. A compressed array commonly requires decompressing and exploring the entire data to perform a filter query. However, in a compressed array in SEACOW, the candidate region for a filter query is selected using the wavelet synopsis and HMMT, and only the chunks in the candidate region are decompressed to perform the query.

However, there are several considerations to be made: (1) the blocks usually have many elements that are close to zero, and (2) each block $B_i$ requires different $R_i^{real}$ bits to encode its elements and has a different distribution of elements according to $R_i^{real}$. First, we observe that the elements in the blocks are generally close to zero and also approach a normal distribution. In the result of the wavelet transform, the significant values are compacted into approximate coefficients; the target blocks for the additional compression contain detailed coefficients, which are less significant values. Our experimental results described here demonstrate a greater probability that the detail coefficients are encoded as small values. Second, we also observe that the block $B_i$ possesses a different distribution of elements depending on the $R_i^{real}$ and this requires a different Huffman coding table.

We experimentally obtain the static Huffman coding table. First, we perform data analysis on our dataset, namely STAR (described in Table 2). The analysis is conducted on the wavelet transformed array. We classify the array blocks according to their $R_i^{real}$ and examine the blocks by counting the number of elements according to their values. Then, we attempt to fit a probability density function (PDF) of normal distribution to the counting data. As a result, we obtain a different PDF according to $R_i^{real}$. Clearly, a block $B_i$ with a high $R_i^{real}$ has a high variance, whereas a block with a low $R_i^{real}$ has a low variance. In every case, the mean of the distribution is always close to zero. For example, we obtain the normal distribution $X \sim \mathcal{N}(\mu=0, \sigma=17.4)$ from the set of elements bit-packed in 7 bits and $X \sim \mathcal{N}(\mu=0, \sigma=35.7)$ from the set of elements bit-packed in 8 bits. Finally, we build a set of static Huffman trees based on given PDFs and use them in the compression process. In the compression process, the blocks adapt different static Huffman coding tables according to $R_i^{real}$, which is the minimum required bits length for bit-packing their elements. As an exceptional case, Huffman coding is not applied when $R_i^{real}$ of a block is zero or one. Note that all datasets use the identical set of Huffman trees obtained from this experiment.

### F. TIME AND SPACE COMPLEXITY ANALYSIS
As mentioned earlier, there are three steps for SEACOW compression. First, in the preprocessing step, we build an HMMT and compress the tree. The time complexity for handling HMMT is $O(N \log N)$, where $N$ is the number of array cells. We also apply the wavelet transform for an array in the preprocessing step, and the time complexity of the transform is $O(N)$. Second, the time complexity of the bit-packing process is $O(N)$. Finally, the time complexity of bit-reduction including run-length encoding and static Huffman coding is $O(N)$. It is noteworthy that there is no cost for building the Huffman tree, as we use the static Huffman coding table, which is pre-built before the compression process. Overall, the time complexity of SEACOW is $O(N \log N)$.

On the other hand, the space complexity of SEACOW is $O(N)$. In particular, we use the lifting scheme for the wavelet transform and the space complexity of the wavelet transform is $O(N)$. The space complexity of both the bit-packing and bit-reduction is also $O(N)$.

## IV. QUERY PROCESSING ON SEACOW
In this section, we present query processing algorithms for SEACOW. In particular, we focus on two types of exploration queries: value-based and dimension-based exploration.

### A. FILTER QUERY PROCESSING
A value-based exploration query is used to select only the values of interest or to find specific data patterns. In this context, we use the term *filter query* to refer to a query that finds a specific element. A simple strategy of the filter query can be achieved by loading the entire array from a disk and examining all elements in the array to locate the specific element. However, this simple strategy would incur high I/O and computational costs to retrieve and decompress the entire array.

However, if the array has an index for the elements, using an index structure can significantly improve the filter query performance. Unfortunately, it is not available in array databases that hold no attribute indices. However, SEACOW embeds a synopsis in its compressed bit-stream, which can subsequently be utilized as a value index. In particular, HMMT helps to find a candidate region for the target value of the query. Algorithm 3 shows how to utilize HMMT to find the candidate region.

---

**Algorithm 3:** Finding a Candidate Region for a Filter Query Using HMMT

**Input** : a predicate $p$ of a filter query, an HMMT **T**, a chunk level $L_{chunk}$, a block level $L_{block}$

**Output:** a chunk-block bitmap **M**

1   Initialize(**M**), $level \leftarrow 0$
2   $\mathbb{N}_{cur} \leftarrow$ GetRootNode(**T**)
3   **while** $\mathbb{N}_{cur} \neq \varnothing$ *and* $level \leq L_{block}$ **do**
4     $\mathbb{N}_{next} \leftarrow \varnothing$
5     **if** $level = L_{chunk}$ **then**
6       **foreach** $N$ *in* $\mathbb{N}_{cur}$ **do**
7         $ChunkId \leftarrow$ GetChunkId($N$)
8         **if** *Evaluate($N$, $p$)* **then**
9           /* Set a chunk bitmap      */
10           **M**[$ChunkId$] $\leftarrow$ true
11           $\mathbb{N}_{next} \leftarrow \mathbb{N}_{next} \cup$ GetChildNodes($N$)
12         **else**
13           **M**[$ChunkId$] $\leftarrow$ false
14         **end**
15       **end**
16     **else if** $level = L_{block}$ **then**
17       **foreach** $N$ *in* $\mathbb{N}_{cur}$ **do**
18         $ChunkId \leftarrow$ GetChunkId($N$)
19         $BlockId \leftarrow$ GetBlockId($N$)
20         **if** *Evaluate($N$, $p$)* **then**
21           /* Set a block bitmap      */
22           **M**[$ChunkId$][$BlockId$] $\leftarrow$ true
23         **else**
24           **M**[$ChunkId$][$BlockId$] $\leftarrow$ false
25         **end**
26       **end**
27     **else**
28       **foreach** $N$ *in* $\mathbb{N}_{cur}$ **do**
29         **if** *Evaluate($N$, $p$)* **then**
30           $\mathbb{N}_{next} \leftarrow \mathbb{N}_{next} \cup$ GetChildNodes($N$)
31         **end**
32       **end**
33     **end**
34     $level \leftarrow level + 1$
35     $\mathbb{N}_{cur} \leftarrow \mathbb{N}_{next}$
36   **end**
37   **return** $\mathbb{M}$

---

The algorithm traverses the nodes of the given HMMT $T$ in breadth-first order. It begins with the root node, which is placed at level zero of $T$; the nodes required for the iteration are listed in $\mathbb{N}_{cur}$. The algorithm evaluates that each node $N$ satisfies the given predicate $p$, where $N$ has a pair of variables $min$ and $max$. The variables in each node indicate the value range of elements in the matching fragment. Accordingly, we can estimate whether $N$ would have elements satisfying $p$ or not. Suppose that $p$ specifies to find the value $v$ in the array. If $v$ is located outside the range [$min$, $max$], there should be no element to find. In this case, there is no further action for the child nodes. On the contrary, if the $v$ is between the [$min$, $max$], we should further explore its child nodes to obtain the more specific region of the element to be found. Using this fact, a function *Evaluate* returns true when $N$

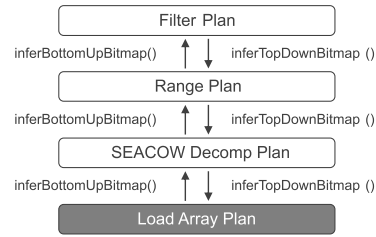

 Example of inferring chunk bitmap with a query plan. The node colored in gray is a leaf node that performs I/O job.

satisfies the condition of $p$, and false otherwise. Meanwhile, there are branches for the two special cases, where the *level* is matched with the chunk level $L_{chunk}$ or the block level $L_{block}$. At each level, the algorithm finds the chunk and block related with $N$ and marks the corresponding bit of the bitmap $M$. As a result of the algorithm, we can obtain the bitmap that ascertains the candidate chunks and blocks. The bitmap helps to eliminate unnecessary chunks and blocks from the scanning procedure and reduces the computation cost as well as the I/O cost.

### B. RANGE QUERY PROCESSING

In this section, we use the term *range query* to refer to a dimension-based exploration query. The range query performance is dependent on the size of data that requires scanning. Improving the range query processing can be performed by excluding the unnecessary chunks located outside the query region. Furthermore, excluding the unnecessary blocks in each chunk reduces the computational cost for decompressing the blocks. However, general compression schemes are not possible because they convert an array into a dummy of compressed bit-stream during the compression process. There is no barrier between the blocks, whereas our scheme supports partial decoding in a chunk.

Most importantly, we should know which part of the array to examine for the query in query planning, for which chunks and blocks that are included or intersected with the query region should be located. The query optimizer examines a query plan tree and infers a candidate region, which is depicted in Figure 7. Inferring the chunk bitmap involves two steps: (1) inspection and (2) propagation. In step (1), the optimizer restricts the query region by referring to the query predicates. This step proceeds from the leaf node (colored in gray) to the root node of the query plan tree. Each node of the query tree passes the chunk bitmap to its parent. For a simple scan query with no boundary, the optimizer marks all the chunks of the array as the region of the array to be used for the parent query. Additionally, for the range query, the optimizer can recognize the exact boundary of the query region according to the given parameters. At the end of step (1), the root has the expected shape of the result array.

However, the I/O job that reads arrays from a disk and un-compresses them is completed on the leaf. Thus, to reduce the I/O cost, the leaves should recognize the restricted region from step (1). In step (2), the optimizer determines the candidate region of the query from the root to the leaf nodes. Each node of the tree restricts the required chunks and blocks according to the chunk bitmap from its parent node. Then, the node conveys the newly created chunk bitmap to its child nodes.

**TABLE 2.** Detailed specifications of eight real-world datasets. In data type column, uint8 and uint16 refer the unsigned integer of one-byte and two-bytes, respectively.

| Name | Data type | Size | Dimensions | Chunk dimensions | Source |
|------|-----------|------|------------|------------------|--------|
| Star | uint8 | 1MB | $1,024 \times 1,024$ | $64 \times 64$ | Hubble Space Telescope, NASA (potw1852a) |
| Jupiter | uint8 | 1MB | $1,024 \times 1,024$ | $64 \times 64$ | Hubble Space Telescope, NASA (heic1914a) |
| Solar | uint8 | 16MB | $4,096 \times 4,096$ | $64 \times 64$ | Solar Dynamics Observatory[2], NASA (HMID 4096) |
| Mars | uint8 | 8MB | $4,096 \times 2,048$ | $128 \times 128$ | USGS Astrogeology Science Center[3] (Viking Colorized Global Mosaic 232m v2) |
| Mercury | uint8 | 128MB | $16,384 \times 8,192$ | $128 \times 128$ | USGS Astrogeology Science Center (MESSENGER Global DEM 665m v2) |
| Lunar | uint16 | 6GB | $98,304 \times 32,768$ | $512 \times 512$ | USGS Astrogeology Science Center (LROC WAC DTM GLD100 118m v1) |
| SDO | uint8 | 0.5GB | $1,024 \times 1,024 \times 512$ | $64 \times 64 \times 64$ | Solar Dynamics Observatory[4], NASA (AIA 171 [2022-01-01 - 2022-01-03]) |
| GridRad | uint8 | 3GB | $1,024 \times 2,048 \times 24 \times 64$ | $64 \times 64 \times 8 \times 8$ | UCAR Research Data Archive[5] (2017-11-01 00:00 - 2017-11-03 15:00) |

## V. EXPERIMENTS

In this section, we present our experimental evaluation. We have two types of proposed methods: SEACOW indicates that the preprocessing (step 1) and the bit-packing (step 2) are applied, and SEACOW+BR indicates that the bit-reduction (step 3) follows SEACOW. We evaluate the compression ratio and query processing performance of the proposed methods against those of existing methods, including COMPASS [6], ZFP [36], Huffman coding [30], LZW (Lempel-Ziv Welch) [28], [29] and SPIHT [19]. Composite compression methods that combine the two existing methods sequentially also exist: LZW+Huffman refers to a compression scheme where Huffman coding follows LZW. In addition, we compare the compression performance with several lossy compression techniques, including SZ (SZ3) [22] and TTHRESH [23]. Both methods enable specifying the target error bounds for their lossy compression. We adjust the error bound to make them a lossless compression. The details are described in the next section.

We employ eight different real-world datasets from a field of astronomy and earth science: Star, Jupiter, Mars, Mercury [50], Lunar [51], and SDO [52], which are image data from astronomy and GridRad, which is weather radar data from NEXRAD WSR-88D weather radars [53]. Table 2 lists the details of the datasets.

We conducted experiments on a Windows machine equipped with an Intel Core i9-11900K (@ 3.50GHz) processor, 128GB of memory, and Samsung 1TB SSD running Windows 10 Enterprise. The experiments were performed on MSDB, our implementation system. As a parallel option, we use six threads for the query processing.

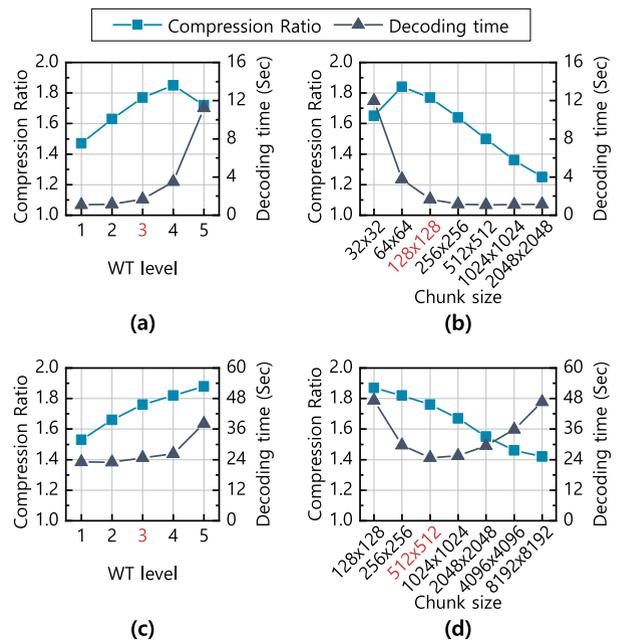

**FIGURE 8.** SEACOW compression performance on Mercury and Lunar data. The results for Mercury are depicted in (a) and (b), and the result for Lunar are depicted in (c) and (d). The parameters used in our experiments are colored in red.

### A. ANALYSIS OF SEACOW COMPRESSION

In this section, we compare the compression performance of SEACOW in terms of user-defined parameters: chunk size and wavelet level. Both parameters are related to compression size and decoding time; however, there are no fixed optimal parameters. Instead, suboptimal parameters can be found experimentally. We performed a comparative study of the two datasets (i.e., Mercury and Lunar) to find suboptimal parameters. In Figure 8, the decoding time includes reading a compressed bit-stream from a disk, decompressing the bit-stream, and reorganizing it as a multidimensional array. First,

**TABLE 3.** Compression ratio comparison. The top three schemes with the best compression ratio are marked in bold. The SZ and TTHRESH are lossy compression schemes, and the brackets indicate that information loss occurs.

| Dataset | COMPASS | SPIHT | ZFP | LZW | Huffman | LZW+ Huffman | SEACOW | SEACOW+ BR | SZ | TTHRESH |
|---------|---------|-------|------|-------|---------|--------------|--------|------------|-------|---------|
| Star | 0.93 | 1.17 | 1.00 | 0.92 | **1.62** | 1.20 | 1.19 | **1.36** | **1.48** | 0.26 |
| Jupiter | 1.41 | **2.83** | 2.18 | 1.93 | 2.38 | 2.27 | **4.13** | **4.57** | 2.68 | (0.74) |
| Solar | 1.09 | 1.59 | 1.41 | 0.99 | 1.68 | 1.30 | **1.91** | **1.96** | **1.71** | (0.45) |
| Mars | 0.80 | 1.27 | 0.98 | 0.91 | **1.33** | 1.08 | 1.27 | **1.54** | **1.32** | 0.38 |
| Mercury | 0.64 | 1.69 | 1.12 | 1.95 | 1.19 | **2.22** | 1.77 | **2.11** | **2.29** | 0.59 |
| Lunar | 0.73 | 1.57 | **2.45** | 1.03 | 1.13 | 1.11 | **1.76** | **1.78** | 1.42 | (0.67) |
| SDO | 0.77 | **2.35** | 1.13 | 1.83 | 1.55 | 1.93 | **2.33** | **2.84** | 1.89 | (0.68) |
| GridRad | 2.14 | 7.66 | 10.45 | **45.98** | 7.39 | **49.94** | 20.44 | 41.21 | **44.34** | (1.84) |

Figures 8(a) and 8(c) present comparison results on varying wavelet levels, while the chunk size is fixed as $128 \times 128$ for Mercury data and $512 \times 512$ for Lunar data. As shown in the results, a higher wavelet level results in a smaller compressed array size and increased decoding time. Additionally, the decoding time starts to increase sharply when the wavelet level is four or higher in both datasets.

Second, Figures 8(b) and 8(d) present comparison results on varying chunk size, while the wavelet level is fixed as 3 in both datasets. In the results, we observe that a smaller chunk size results in decreased storage requirements, but increased decoding time. Nonetheless, a very tiny chunk size could degrade the compression ratio as it generates many small chunks, all of which have individual metadata that require additional storage space. Therefore, using a tiny chunk would be inefficient for both compression ratio and query performance. As the compression ratio and the query performance have a trade-off, the chunk size and the wavelet level should be balanced considering the compressed size and query performance. Through these experiments, we find efficient parameters for SEACOW compression. In the remainder of the experiments, we used *wavelet_levels*=3 and various chunk sizes from $64 \times 64$ to $512 \times 512$ depending on the data size.

In SEACOW, the compressed array can be divided into three parts: HMMT, Synopsis array, and array body. Each of the parts is compressed and stored; in particular, the size of HMMT is significantly reduced through the encoding process presented in Section III-D3. For example, the compression ratios of HMMT in Mercury and Lunar are 4.03 and 6.48, respectively. Accordingly, there are only a few MB in size, small enough to fit in the cache memory. Figure 9 shows the proportion of each part (HMMT, synopsis array, and body) of the compressed array size.

### B. COMPARISON OF COMPRESSION PERFORMANCE

Table 3 reports the compression ratio on our datasets using eight different compression schemes, which are the lossless methods. The result includes all components of the compressed array such as the HMMT and synopsis array. In particular, we set the different error bounds for the lossy compression methods. Setting the error bound to zero is a simple way to make lossless compression; however, neither method works properly with this setting. Instead, we set the *absolute_error* = 0.99 for SZ as we only use integer data. Meanwhile, TTHRESH supports three different error-bound types, and they are converted to the sum of squared error (SSE). In this case, even with a very low error bound, there is some information loss. Furthermore, as the error bound decreases, the size of the compressed array by TTHRESH becomes bigger than the raw array. The compression ratio of TTHRESH in the Table 3 is the execution result when the *relative_error* = 0.00001.

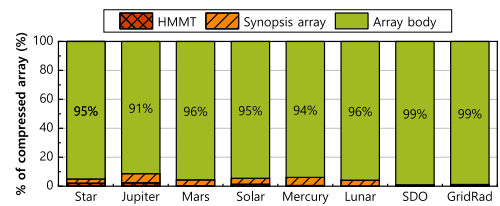

**FIGURE 9.** Percentage of each part (HMMT, Synopsis array, and array body) of a SEACOW compressed array. In the compressed array size, the HMMT and synopsis array only take up 1.46% and 5.01% on average.

In this experiment, SEACOW+BR achieves the highest compression ratio on average. It is noteworthy that the succeeding Huffman coding improves the compression performance by 14% on average, and up to 21% in SEACOW. In general, SEACOW shows a high compression ratio for a dataset with low data distribution; however, it is not the only condition for efficient compression in SEACOW. More specifically, good performance can be achieved when similar values are located adjacent to each other. As an array is stored in the unit of the chunk, the data distribution is valid in the chunk. For example, the Jupiter dataset has a 2.4 times higher standard deviation (STD) than the Star dataset. Reversely, the chunks of Star have a higher STD than those of Jupiter. As a result, in SEACOW, Jupiter shows a higher compression ratio than Star. In the case of GridRad, it has a very low STD, while most of the data is empty with zero. This feature of data is advantageous for compression and leads to a high compression ratio in all compression schemes.

Meanwhile, COMPASS is inefficient in compressing an integer array in our experiments. COMPASS requires additional storage space to note coordinates, in contrast to an array that implicitly represents the position of each array cell. As a result, COMPASS hardly reduces the volume of data in most of our datasets. Accordingly, we excluded COMPASS from the rest of the experiments.

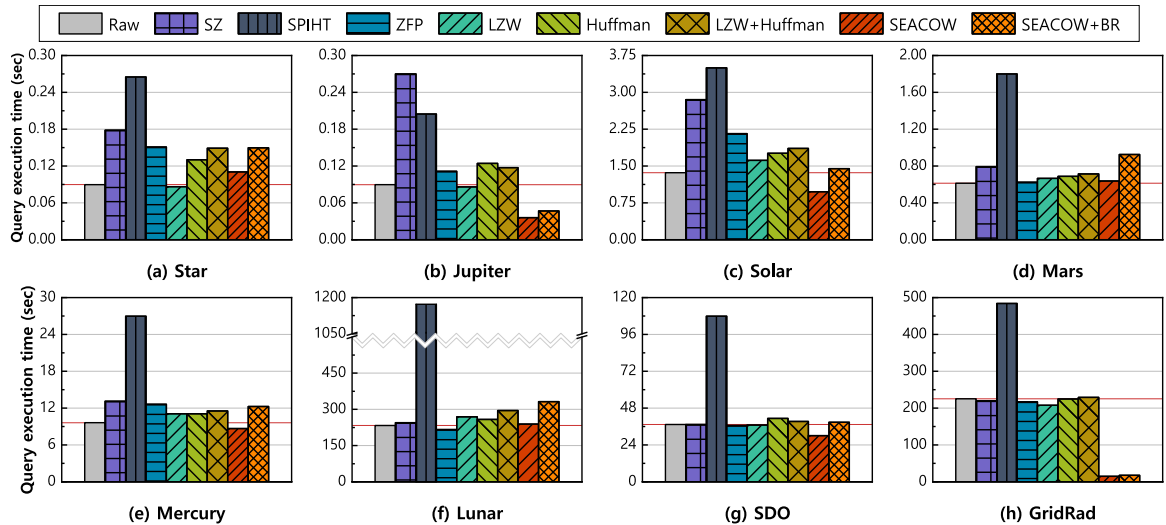

**FIGURE 10.** Filter query performance on various datasets. The red colored line in each graph shows the execution time of Raw as a baseline.

Here, we briefly report the end-to-end compression time, which includes loading data from a file, array partitioning, data encoding, and writing compressed data in a file. As we mentioned earlier, we used six threads for the compression process. In our experiments, SEACOW takes 15% more time on average to compress the data than Raw. SEACOW+BR requires an additional compression process, thereby consuming 18% more time than Raw. For example, it takes 295.7 sec for Raw, 344.6 sec for SEACOW, and 361 sec for SEACOW+BR to encode 6 GB of Lunar data. In addition, SEACOW and SEACOW+BR took 15.5 sec more to build HMMT, where the HMMT building process involves only a single thread. Then, the end-to-end compression time for SEACOW and SEACOW+BR are 360.1 sec and 376.5 sec, respectively.

### C. VALUE-BASED EXPLORATION PERFORMANCE

This section presents the value-based exploration performance, which is finding specific values in an array. The filter query refers to the value-based exploration, which is an essential workflow that is frequently used in data analysis. The filter query leads to a high computational cost to iterate all the array cells and compare each element. The Raw, uncompressed storage method, can start the filtering process directly on the chunks loaded from a disk. In contrast, with compressed techniques, additional computational tasks are required for the decoding process, and generally, query performance might be declined. Figure 10 presents the performance of the filter query on various datasets. We randomly select the three values for the filter query and reported the average query execution time for three runs.

In this experiment, SEACOW shows fast query performance in most datasets, and there is no significant performance degradation compared to Raw. In particular, SEACOW greatly outperformed in the GridRad (in Figure 10(h)), where most of the array cells are empty by zero. It also shows the best performance on Jupiter (in Figure 10(b)), where similar elements are gathered from each other. In these datasets, SEACOW efficiently excludes unnecessary regions of the array using the embedded index structure, HMMT. In Star (in Figure 10(a)), LZW was the fastest, followed by Raw and SEACOW. The Star is an image data that consists of a dark background with numerous randomly located stars, where the background has a lower color value ($0 \times 00$) and the stars have a higher color value (0xFF). For SEACOW, it is the worst case to process filter queries. Because all blocks have elements from 0 to 255, it is not possible to decrease the search region using HMMT. In this case, SEACOW should decode the entire array and perform a filter query on them similar to the other methods.

Conversely, SZ and SPIHT, which cause high computational costs, show poor performance. Particularly, SPIHT incurs huge random access with a computational-intensive decoding algorithm, and is slowest in most datasets. In addition, SEACOW+BR has an additional computational cost compared to SEACOW. Accordingly, SEACOW+BR is 36% slower on average than SEACOW. LZW+Huffman is also 20% slower on average than LZW. Note that, most of the compression schemes achieve a high compression ratio on GridRad, and show comparable query performance with Raw in Figure 10(h).

### D. DIMENSION-BASED EXPLORATION PERFORMANCE

This section presents the query performance of a dimension-based exploration with a filter query finding values in a specific region. The range query refers to the dimension-based exploration, and the range-filter query refers to the combination of the dimension and value-based exploration. Because the multidimensional data is large, it is expensive and inefficient to involve the entire array in an analysis query. Instead, it is common to designate the region of interest and perform analysis only in that region. The selectivity of range querying refers to the percentage of query regions in the entire array space.

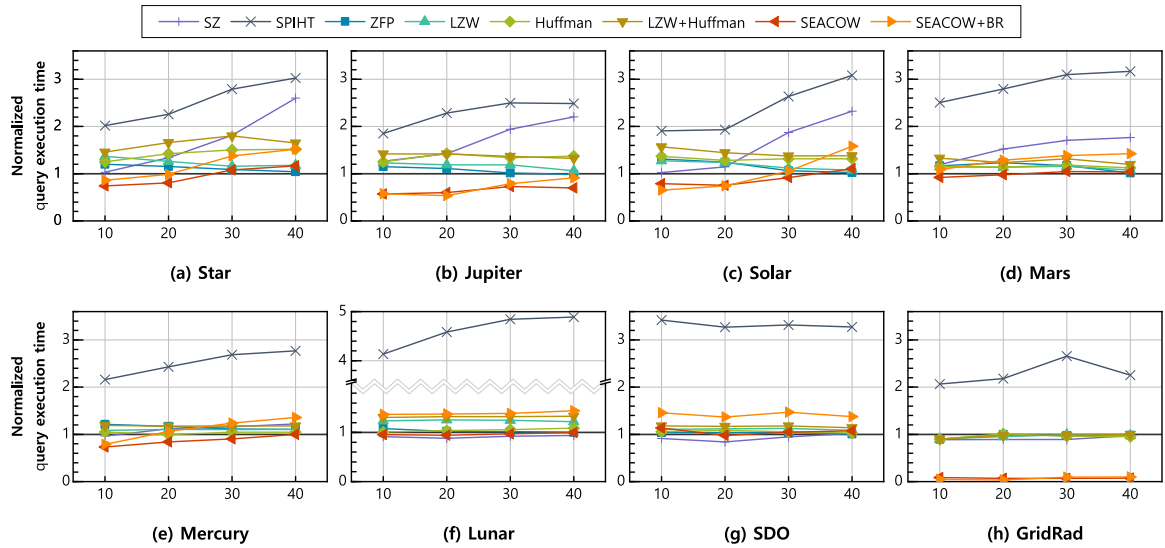

**FIGURE 11.** Filter with range selection query performance on various datasets. The results are normalized based on the query execution time of Raw. The bold horizontal line in each graph is the baseline. The results located below the baseline indicate that the query execution time is faster than Raw, and conversely, results located above the line indicate the query execution time is slower than Raw.

Figure 11 shows the range-filter query execution time on various datasets. The x-axis indicates the selectivity of the range query. It is varied from 10(%) to 40(%). We generated three random regions for each selectivity and reported the average query execution time for these queries.

In general, the compressed arrays took 10%-30% more time to perform the range query than the uncompressed array. Particularly, SPIHT is 2.8 times slower than Raw. Nonetheless, SEACOW shows almost similar performance compared to Raw, and in some cases, outperformed it. In particular, SEACOW is faster than Raw on Jupiter (in Figure 11(b)) and GridRad (in Figure 11(h)), which have high compression rates. Particularly, in GridRad, SEACOW can greatly reduce the number of chunks involved in query processing by utilizing HMMT.

In this experiment, the query performance is affected by the size of the querying region. The query execution time increases as the selectivity increases, though it is not represented in the normalized graph. On the other hand, the computationally intensive compression schemes, such as SZ, SPIHT, and SEACOE+Huffman, show relatively poor query performance in the higher selectivity queries than the lower ones. It might indicate that the computational cost is more significant than the I/O cost in queries with high selectivity. Note that, in Figure 11(a), (c), and (e), SEACOW+BR still outperforms Raw in queries with a small region. SEACOW and SEACOW+BR take the advantage of partial decoding in lower selectivity queries. The partial decoding might be useful in chunks located at the edge of the query range.

In these chunks, SEACOW performs the decoding process in the unit of blocks according to the query range. As a result, it performs better with small query regions on Star and Jupiter (in Figure 11(a) and (b)), which are the smallest datasets.

## VI. CONCLUSION

We presented a lossless compression algorithm called SEACOW for multidimensional arrays. The proposed method employs a wavelet transform that is widely used in multimedia file compression. In fact, it demonstrates as high compression ratio compared to those of well-known existing compression algorithms in our experiments. It is noteworthy that the key difference from the existing methods is that SEACOW embeds additional data structures called the synopsis array and HMMT in a compressed array. These data structures are used in the compression process. Furthermore, they can accelerate exploration query processing. Utilizing the feature of the HMMT, SEACOW also shows good performance in exploration query processing. As a result, the proposed method offers a good balance of the compression ratio and exploration query performance. There are various applications that can benefit from improved performance by utilizing the synopsis, such as the calculation of an aggregate on each non-overlapped sub-array or the determination of Top-K values in a specific region. A limitation of our study is that SEACOW allows only the integer type of numbers except for the floating-point values. Instead, many real-world data with fixed point precision can be easily converted into an integer variable by multiplying them by a specific value $10^{pr}$, where $pr$ is the precision. Then, they can be treated the same as integer data in SEACOW.

In future work, supporting distributed processing is required to process a massive volume of data. In our study, each array is partitioned into chunks and can be extended to a distributed storage environment. In particular, the synopsis could be replicated and stored on a master node to be utilized in the query optimizer. In addition, we will apply progressive query processing that could be useful for various exploration tasks in SEACOW.

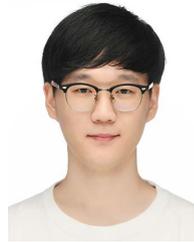

**HYUBJIN LEE** received the B.S. degree in computer science from Korea University, Seoul, South Korea, in 2020, where he is currently pursuing the M.S. degree with the Department of Computer Science and Engineering. His research interests include array database, analytical processing, data privacy, and machine learning.

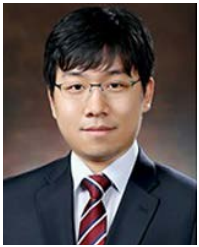

**MINSOO KIM** received the B.S. degree in computer and information science from Korea University, Sejong, South Korea, in 2015, where he is currently pursuing the Ph.D. degree with the Department of Computer Science and Engineering. His research interests include array database and distributed/parallel processing of large-scale data.

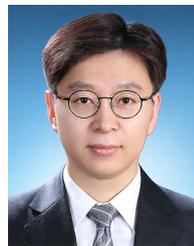

**YON DOHN CHUNG** (Member, IEEE) received the B.S. degree in computer science from Korea University, Seoul, South Korea, in 1994, and the M.S. and Ph.D. degrees in computer science from KAIST, Daejeon, South Korea, in 1996 and 2000, respectively. He was an Assistant Professor with the Department of Computer Engineering, Dongguk University, Seoul, from 2003 to 2006. He joined as the Faculty Member of the Department of Computer Science and Engineering, Korea University, in 2006, where he is currently a Professor. His research interests include array database, distributed/parallel processing of large-scale data, spatial databases, and data privacy.

• • •